\documentclass[conference]{IEEEtran}

\IEEEoverridecommandlockouts
\usepackage{amsmath}
\usepackage{color}
\usepackage{array}
\usepackage{stfloats}
\usepackage{amsthm} 
\usepackage{amssymb}
\usepackage{float}
\usepackage{nccmath}
\usepackage{graphicx}
\usepackage{algorithm}
\usepackage{algorithmic}
\usepackage{setspace}
\usepackage{booktabs}
\usepackage{subcaption}
\usepackage{mwe}
\usepackage{cite}
\usepackage{amsmath,amssymb,amsfonts}
\usepackage{graphicx}
\usepackage{textcomp}
\usepackage{xcolor}
\usepackage[utf8]{inputenc} 
\usepackage{kotex} 

\usepackage{xspace}
\def\BibTeX{{\rm B\kern-.05em{\sc i\kern-.025em b}\kern-.08em
    T\kern-.1667em\lower.7ex\hbox{E}\kern-.125emX}}
\begin{document}
\newcommand{\BfPara}[1]{{\noindent\bf#1.}\xspace}
\newcommand\mycaption[2]{\caption{#1\newline\small#2}}
\newcommand\mycap[3]{\caption{#1\newline\small#2\newline\small#3}}

\title{Distributed and Autonomous Aerial Data Collection in Smart City Surveillance Applications}

\author{
\IEEEauthorblockN{Haemin Lee, Soyi Jung, and Joongheon Kim}
\IEEEauthorblockA{Department of Electrical and Computer Engineering, Korea University, Seoul, Republic of Korea \\
E-mails: \texttt{haemin2@korea.ac.kr}, \texttt{jungsoyi@korea.ac.kr}, \texttt{joongheon@korea.ac.kr}}
}

\maketitle

\begin{abstract}
The massive growth of Smart City and Internet of Things applications enables safety and security. The data those are produced from surveillance cameras in aerial devices such as unmanned aerial networks (UAVs) are needed to be transferred to ground stations for secure data analysis. When the scale of network is relatively large compare to the wireless communication coverage of device, it is not always available to transmit the data to the ground stations, thus distributed and autonomous algorithms are essentially desired. Based on the needs, we propose a novel algorithm that is for collecting surveillance data under the consideration of mobility and flexibility of UAV networks. Due to the battery limitation in UAVs, we selectively collect data from the UAVs by setting rules under the consideration of distance and similarity. As a sequence, the UAV devices have to compete for a chance to get data processing. For this purpose, this paper designs a Myerson auction-based deep learning algorithm to leverage the UAV's revenue compare to traditional second-price auction while preserving truthfulness. Based on simulation results, we verify that our proposed algorithm achieves desired performance improvements.
\end{abstract}

\section{Introduction}\label{sec:1}
Urban areas composed with the high density of population and traffic suffer from safety issues and transport efficiency~\cite{KIM2017159,9143155}. As a prevention of threats, various sensing and monitoring Internet of things (IoT) devices have been deployed throughout the city-wide areas~\cite{isj2021dao}. Thus, it is obvious that IoT devices make the city smarter and secure. Closed circuit television (CCTV) is one of the main supervision tools in smart city surveillance applications and deployed in Point-of-Interest (POI), that is, geometrically and socially important spots or crime prone areas. Furthermore, the CCTV-recorded data is real-time video streaming, thus the corresponding streaming and scheduling technologies are also actively discussed~\cite{ton201608kim,jsac201806choi,tmc201907koo,twc201912choi,tmc2021yi,mm2017koo}. The real time monitoring video data from CCTV enable facility infrastructure management, spatial information acquisition, and adequate response to rising problems in smart city applications.

Most of monitoring/sensing data are a time-sensitive deadline, and it is mostly valid for a specific period of time. In this paper, we propose a strategic data collection from sparsely located CCTV devices. We assume the situation when the intelligent IoT CCTV cannot transmit or relay sensing data to a base station via multi-hop relays. Instead, we consider the case where we deploy unmanned aerial vehicles (UAVs) to selectively collect data and planning the flying route~\cite{tvt202106jung,tvt2021jung}. UAV provides more extensive and diverse application with its increase of utilization~\cite{9406452,electronics20jung,10.4108/eai.30-6-2020.165502}. Due to the natural trait of UAVs, the UAVs can swiftly move and optimize their path in order to quickly complete their mission~\cite{access19geraldes}. UAVs can also collect raw data and deliver it to the target point operating. Furthermore, the communication between a ground device to a UAV in the air has an advantage in radio signal degradation, unlike how the signals degrade quickly for the wireless communication between two devices on the ground due to various shadowing and scattering~\cite{chandrasekharan2016}. 

However, due to the battery limitations of UAVs, it is occasionally unrealistic to collect all data in POIs. Instead, we take an economic approach in order to selectively collect data under the consideration of distance and data similarity. In this paper, we design a novel algorithm that is for the data collection with Myerson auction approach for distributed and autonomous data collection using UAVs. Furthermore, we utilizes deep learning-based framework for solving the Myerson auction-based formulation for optimizing seller's revenue. 
There are several application researches to solve resource allocation problems with the variant Myerson auctions~\cite{shin2019auction, luong2018, dutting2019optimal}.
The contribution of this paper is two folds. First, we collect the data from CCTV efficiently in terms of distance and data redundancy. In addition, we maximize the seller's revenue with deep learning auction approach. 

The rest of the paper is organized as follows. 
Sec.~\ref{sec:2} and Sec.~\ref{sec:3} propose the system model and our auction model, respectively. Sec.~\ref{sec:4} evaluates the performances and Sec.~\ref{sec:5} concludes the paper.

\section{Data Collection System Model}\label{sec:2}
\subsection{Overall Architecture}\label{sec:2-1}
\begin{figure}[t!]
    \centering
    \includegraphics[width=1.0\columnwidth]{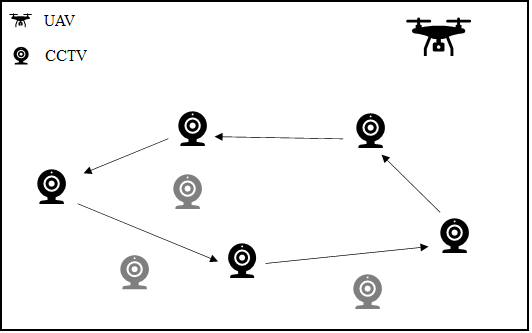}
    \caption{Data collection scenario by UAV.}
    \label{fig:overview}
\end{figure}

This section introduces the overall architecture of our data collection scenario and auction process. The proposed system consists of a UAV and sparsely located CCTV devices installed in the POIs throughout the city as shown in Figure~\ref{fig:overview}. Due to the relatively long distance between the devices, sometimes it is unrealistic for them to transmit or relay sensing data to a base station.
We assume that the UAV can only assigned to a single device at a time. As a result, the devices compete for the drone's data processing ability to transmit their collected data.

The UAV acts as a seller and the devices act as a buyer.
The devices in need strategically submit their bid based on their valuation.
Then the UAV collects all bids which are in transform form. The drone with the highest allocation probability becomes a winner and pays final the determined payment and is detailed in Sec.~\ref{sec:3}.
Through the sequence of auctions, the trajectory of UAV is determined.

\subsection{Private Valuation}\label{sec:2-2}
Note that each device $u_i$ have their own private valuation $v_i$ as equation~\ref{eq:valuation}. 
We assume that every bidder devices know the last collected data and location of each drone. As the distance between the device and the drone gets close, the device would have a higher winning probability which means it is willing to attend the auction. 

Devices are located in POI, while the location of the drone varies. The distance $d_i$ between the drone and the device can be derived as
\begin{equation}
    d_i = \sqrt{{(x_{u}(t) - x_{i})^2}+{(y_{u}(t) - y_{i})^2}},
    \label{eq:euclidean}
\end{equation}
where $x_{u}(t)$ and $y_{u}(t)$ denote the location of UAV at time t, and $x_{i}$ and $u_{i}$ denote the location of device $i$. In addition, if the current bidder's data is similar to the data collected from just before round, it means the data may be redundant. The device with high similarity, thy would be less active in the auction.
For all images in the pile, we one by one compare two $m$ $\times$ $n$ sized images for similarity $s_i$, using mean squared error (MSE) for all pixels. 
$K = \{1,2,..,|K|\}$ denotes the image pile collected just before round, and $I = \{1,2,..,|I|\}$ denotes the current device's image pile, and the MSE (denoted as \textsf{MSE} below) can be derived as follows,
\begin{equation}
    s_i = \textsf{MSE} = \frac{1}{mn} \sum_{i = 0}^{m-1}\sum_{i = 0}^{n-1}[{I(i,j) - K(i,j)]^2}.
    \label{eq:similarity}
\end{equation}

In general, when the distance $d_i$ and data similarity $s_i$ are large, the buyer would be less incentive. Therefore, the valuation $v_i$ can be expressed by
\begin{equation}
    v_i = s_i \cdot d_i.
    \label{eq:valuation}
\end{equation}

\section{Algorithm Design Concepts}\label{sec:3}
Myerson presents provable analytical results for single item auctions optimizing the auctioneer revenue in truthful settings where each buyer has its own private valuation of the resource \cite{myerson1981optimal}. 
For a single-item auction with $N$ bidders, Myerson’s mechanism firstly introduces a function of bidder's valuation which is known as the virtual valuation \cite{chawla2007algorithmic} as in~\eqref{eq:valuation},
\begin{equation}
    \phi_i(v_i) = v_i - \frac {1 - F_i(v_i)} {f_i(v_i)}.
\end{equation}

Each bidder $i$ has its own individual private valuation $v_i$ which is drawn from the cumulative density function $F_i(v_i)$ where the probability density function of $v_i$ is defined as $f_i(v_i)$. 
With the concept of virtual valuation, the winner and the final payment are determined. The winner would be the one with the highest virtual valuation. The final payment $q_i$ can be calculated through the second-highest virtual valuation of the user using~\eqref{eq:payment}. This means that a winning bidder pays a price equal to the virtual valuation-inverse of the second-highest virtual valuation, 
\begin{equation}
    q_i = \phi_i^{-1}(\phi_j(v_j)).
    \label{eq:payment}
\end{equation}

We now show how the Myerson variant deep learning-based auction maximizes the expected revenue of UAV while guaranteeing truthfulness and revenue-optimal. Detailed neural architectures for deep learning to solve our proposed auction-based problems are organized in Algorithm~\ref{al:deep}. The monotonic network takes the role of virtual valuation in Myerson auction~\cite{luong2018}. The allocation network maps the UAV and the device with the highest non-zero transform bid. The payment network determines the final payment to the winner delivery UAV.
The neural architecture parameters $w^{i}_{kj}$ and $\beta^{i}_{kj}$ are trained with the valuation profiles as the training set while minimizing the loss function.

\begin{algorithm}[t]
   \caption{Deep Learning-Based Auction Algorithm}
   \label{alg:deep}
\begin{algorithmic}[0]
   \STATE {\bfseries Input:} Candidate bid sets $\textbf{b}=(b_{1}, b_{2},...,b_{N})$. \\
   \STATE {\bfseries Output:} Allocation probability set $g_i=(g_{1}, g_{2},...,g_{N})$, payment set $p_i=(p_{1}, p_{2},...,p_{N})$.
   \REPEAT 
   \STATE Compute \ $\phi_i(b_i) = \max_{\forall k \in K}\min_{\forall j \in J} \left(w^{i}_{kj}b_i + \beta^{i}_{kj}\right)$ ; 
   \STATE Compute \ $g_i(\bar{b}) = \frac{e^{k\bar{b}_i}}{\sum_{j=1}^{N+1}e^{k\bar{b}_j}}$ ;
   \STATE Compute $p_i^{0}(\bar{b}) = ReLU(\max_{\forall j \neq i}\bar{b_j})$ ;
   \STATE Compute\ $\phi^{-1}_i(y) = \min_{\forall k \in K}\max_{\forall j \in J} (w^{i}_{kj})^{-1}\left(y-\beta^{i}_{kj}\right)$;
   \STATE Compute \ $L(w,\beta) =-\sum_{i=1}^{N}g_i^{(w,\beta)}(v^{s})p_i^{(w, \beta)}(v^{s})$ ;
   \UNTIL{The loss function $L(w,\beta)$ converges to the minimum.}
\end{algorithmic}
\label{al:deep}
\end{algorithm}

\section{Performance Evaluation}\label{sec:4}
\begin{figure}[t!]
    \centering
    \includegraphics[width=1.0\columnwidth]{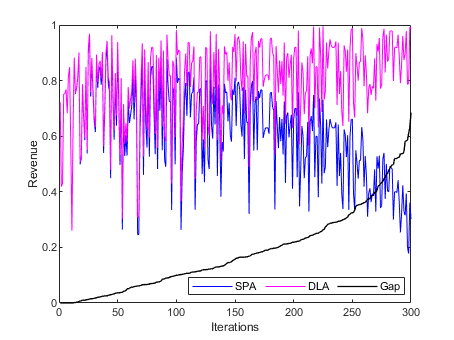}
    \caption{Revenue gap from 300 experiment cases sorted in an scending order.}
    \label{fig:revenue gap}
\end{figure}

In this section, we have a deep learning-based optimal auction (DLA) algorithm for data collection. In addition, the proposed deep learning based optimal auction compared with SPA as a baseline.
We construct a neural network with PyTorch library. To evaluate our system, we performe the deep learning auction where the numbers of participating devices are five with distribution of valuation $f_{V}(v)$ $\sim U[0.5,1]$. We set the five groups and three linear functions for the neural network. Overall 500 iterations were done for every round.
Fig. \ref{fig:revenue gap} shows the 300 individual deep learning auction results. The revenue gap between SPA and DLA is obtained for each iteration, and we sort it in an ascending order. That is, the corresponding graph shows the range of gaps that can occur over iterations. Overall, with DLA, we can confirm the revenue is improved compare to SPA.

\section{Concluding Remarks}\label{sec:5}
In this paper, we propose a distributed and autonomous aerial data collection in smart city surveillance applications. We collect the data from CCTV device selectively, under the consideration of distance and data redundancy. With the deep-learning auction, our scenario achieves the initial objective that is the maximization of revenue of the seller UAV as well as the preserving the truthful conditions in distributed resource allocation. The evaluation results confirm that the auction-based resource allocation formulation between the CCTV devices and UAV gives distinct revenue benefits compared to the traditional SPA.

\section*{Acknowledgment}

This work was supported by the National Research Foundation of Korea (NRF) grant funded by the Korea government (MSIT) (2019M3E4A1080391, 2021R1A4A1030775). S. Jung and J. Kim are the corresponding authors of this paper (e-mails: jungsoyi@korea.ac.kr, joongheon@korea.ac.kr).

\bibliographystyle{IEEEtran}
\bibliography{ref,ref_aimlab}

\end{document}